*Perspective*

# A Roadmap for Improving Data Reliability and Sharing in Crosslinking Mass Spectrometry

Juri Rappsilber[1#], James Bruce[2], Colin Combe[1], Stephen Fried[13], Albert J R Heck[3], Claudio Iacobucci[4], Alexander Leitner[5], Karl Mechtler[6], Petr Novak[7], Francis O'Reilly[8], David C. Schriemer[9], Andrea Sinz[10], Florian Stengel[11], Andrea Graziadei[14], Konstantinos Thalassinos[12]

1 Technische Universität Berlin, Chair of Bioanalytics, 10623 Berlin, Germany; Wellcome Centre for Cell Biology, University of Edinburgh, Edinburgh EH9 3BF, UK; Si-M/""Der Simulierte Mensch"", a Science Framework of Technische Universität Berlin and Charité - Universitätsmedizin Berlin, Berlin, Germany.

2 Department of Genome Sciences, University of Washington, Seattle, WA, USA

3 Biomolecular Mass Spectrometry and Proteomics, Bijvoet Centre for Biomolecular Research and Utrecht Institute for Pharmaceutical Sciences, Utrecht University, Utrecht, Netherlands.

4 Department of Physical and Chemical Sciences, University of L'Aquila, Via Vetoio 67100 L'Aquila, Italy

5 Institute of Molecular Systems Biology, Department of Biology, ETH Zürich, 8093 Zurich, Switzerland

6 Research Institute of Molecular Pathology (IMP), Vienna BioCenter, Vienna, Austria; Institute of Molecular Biotechnology (IMBA), Austrian Academy of Sciences, Vienna BioCenter, Vienna, Austria; Gregor Mendel Institute of Molecular Plant Biology (GMI), Austrian Academy of Sciences, Vienna BioCenter, Vienna, Austria

7 Department of Biochemistry, Faculty of Science, Charles University, Prague, Czech Republic; Institute of Microbiology, The Czech Academy of Sciences, Vestec, Czech Republic.

8 Center for Structural Biology, Center for Cancer Research, National Cancer Institute, National Institutes of Health, Frederick, MD, USA.

9 Department of Biochemistry and Molecular Biology, University of Calgary, AB, Canada, T2N 4N1

10 Center for Structural MS, Martin Luther University Halle-Wittenberg

11 University of Konstanz, Department of Biology, Universitätsstrasse 10, 78457 Konstanz, Germany; Konstanz Research School Chemical Biology, University of Konstanz

12 Institute of Structural and Molecular Biology, Division of Biosciences, University College London, London, WC1E 6BT, United Kingdom; Institute of Structural and Molecular Biology, Birkbeck College, University of London, London, WC1E 7HX, United Kingdom

13 Department of Chemistry, Johns Hopkins University, Baltimore, Maryland 21218, United States; T. C. Jenkins Department of Biophysics, Johns Hopkins University, Baltimore, Maryland 21218, United States.

14 Human Technopole, Via Rita Levi Montalcini 1, 20157, Milano, Italy

# communication to Juri Rappsilber, juri.rappsilber@tu-berlin.de

**Abstract**

Crosslinking Mass Spectrometry (MS) can uncover protein–protein interactions and provide structural information on proteins in their native cellular environments. Despite its promise, the field remains hampered by inconsistent data formats, variable approaches to error control, and insufficient interoperability with global data repositories. Recent advances, especially in false discovery rate (FDR) models and pipeline benchmarking, show that Crosslinking MS data can reach a reliability that matches the demand of integrative structural biology. To drive meaningful progress, however, the community must agree on error estimation, open data formats, and streamlined repository submissions. This perspective highlights these challenges, clarifies remaining barriers, and frames practical next steps. Successful field harmonisation will enhance the acceptance of Crosslinking MS in the broader biological community and is critical for the dependability of the data, no matter where it is produced.

**Highlights**

- Prerequisites for standardisation and data sharing in Crosslinking MS have been fulfilled
- A federated repository ecosystem for Crosslinking MS data is both vital and achievable
- Robust error models will broaden Crosslinking MS data reach and impact
- Agreed benchmarking procedures are needed to ensure data reliability across the field

**In Brief**

Crosslinking Mass Spectrometry (MS) can characterize the topology of proteins and their interactions. However, fragmentation in data standards, repository submission processes, and error control methods limits its potential. This perspective proposes solutions built around mzIdentML 1.3, FAIR data sharing through PRIDE, and deeper integration with PDB-IHM and UniProt. Together, these improvements set the stage for a federated ecosystem that advances Crosslinking MS beyond the FAIR principles – ensuring data quality, facilitating collaboration, and embedding Crosslinking MS firmly within the structural biology toolkit.

## Introduction

Protein–protein interactions (PPIs) underpin much of cellular biology, shaping macromolecular assemblies and regulatory networks. Crosslinking MS addresses these complexities by covalently linking residues in close proximity, capturing distance restraints that can be identified through mass spectrometry (1–6). This fills a gap where classical structural techniques sometimes fail, either due to challenging sample requirements, flexible proteins or transient interactions.

Despite substantial progress in instrumentation, crosslinker chemistry, and computational pipelines, the field has been slow to consolidate practices for data exchange and interpretation (7, 8). Unlike X-ray crystallography or cryo-EM – where community standards, validation procedures, and shared databases are well-established – Crosslinking MS still grapples with inconsistencies in reported data, insufficient provenance tracking, and patchy compatibility with major scientific data repositories. Beyond binary interaction mapping, quantitative Crosslinking MS is now revealing conformational states through co-varying crosslinks (9–11), underscoring the need for formats and repositories to support quantitative data.

At the 2024 Symposium for Structural Proteomics (SSP) in Cambridge, Massachusetts, we asked the community for their views on adopting mzIdentML as the field standard for Crosslinking MS data. During the technical session on data sharing and error control in Crosslinking MS, 78 attendees participated in a live poll consisting of four questions, each with an agreement scale from 1 (low) to 5 (high): (1) "Standardisation is important for the crosslinking MS field": average 4.9, median 5; (2) "mzIdentML and ProteomeXchange resources such as PRIDE represent the way to go": average 4.2, median 5; (3) "Accuracy is important for the crosslinking MS field": average 5.0, median 5; (4) "Crosslinking MS can, in fact, be very accurate": average 4.5, median 5.

In this perspective, we chart a path to remove current barriers. We focus on three imperatives: finalising consistent file formats for reporting crosslink identifications; implementing robust and transparent false discovery rate (FDR) models and standards; and establishing a federated repository network that embraces Crosslinking MS data alongside other structural and functional annotations. Together, these steps would mandate high standards for Crosslinking MS data and dramatically increase the utility, accessibility, and visibility of the data.

## Developing Standardised File Formats

One of the field's longstanding challenges has been the absence of a standardised format for representing crosslinked peptide-spectrum matches. Early efforts typically consisted of spreadsheet-style tables embedded in supplementary materials, with each research group opting for its own layout. This fragmentation hindered downstream interpretation, reproducibility, and cross-lab data reuse.

In response, the HUPO-Proteomics Standards Initiative (HUPO-PSI) extended the mzIdentML standard (12) to cover Crosslinking MS (13), culminating in mzIdentML 1.3.0 (14). This version enables encoding cleavable crosslinkers, multiple spectra assignments, non-covalently associated peptides (15), and internally linked peptides. By capturing sufficient metadata for each identification, mzIdentML 1.3.0 supports data re-analysis, fosters transparency, and paves the way for deeper integration with structural biology pipelines.

For the community at large, adopting mzIdentML 1.3.0 means that any repository or visualisation platform can retrieve, visualise, and validate crosslink identifications in a uniform, machine-readable manner. Specific database search tools such as SCOUT and Kojak (16, 17), and general tools for error estimation that is in principle independent of the search software, xiFDR (18) are already adhering to this standard, offering exporting functionalities. In addition, the data visualisation suite xiVIEW (19) can read mzIdentML and offers interactive visualisation of protein interaction networks (20) and spectra (21) within a web browser. The field's commitment to data standards builds on an established foundation, as numerous tools have already adopted mzIdentML for writing and/or reading crosslink data (Table 1).

**Table 1.** Software tools and databases that support mzIdentML at the time of writing.

| Software | Functionality | mzIdentML support |
|---|---|---|
| xiFDR (for xiSEARCH) (18, 22, 23) | MS data analysis | Exports search results as mzIdentML v.1.3 |
| SCOUT (17) | | Exports search results as mzIdentML v.1.3 |
| Kojak (16, 24) | | Exports search results as mzIdentML v.1.3 |
| Mascot (25) | | Exports search results as mzIdentML v.1.2 |
| ProteomeDiscoverer (26) | | Exports search results as mzIdentML v.1.2 |
| ProteinProspector (27, 28) | | *In progress* |
| MeroX (29) | | *In progress* |
| IMP (30) | Modelling | *Import of restraints from mzIdentML in work* |
| AlphaLink (30, 31) | | *Import of restraints from mzIdentML in work* |
| xiVIEW (19) | Visualisation | Visualises Crosslinking MS data from mzIdentML |
| IHMValidation | Validation | Imports restraints from mzIdentML |
| python-IHM ([https://python-ihm.readthedocs.io/en/stable/index.html](https://python-ihm.readthedocs.io/en/stable/index.html)) | Curation | Imports restraints from mzIdentML |
| PRIDE (32) | Database | Archives Crosslinking MS data in mzIdentML |
| PDB-IHM (33) | | Validates structure models based on crosslinks in mzIdentML format held in PRIDE |

## FAIR Data Sharing

Beyond a common file format, Crosslinking MS must align with FAIR (Findable, Accessible, Interoperable, Reusable) data sharing principles (7, 34). In proteomics, PRIDE (the core partner of the ProteomeXchange consortium (35)) serves as the principal global archive (36). Although PRIDE has accepted Crosslinking MS data previously, these submissions often lacked critical metadata and unified data structures. By adopting mzIdentML 1.3.0, researchers can now deposit the results of their Crosslinking MS experiments (including raw files and search results) in a machine-readable format.

Integration with PRIDE also opens the door to automated data flow. PRIDE's robust Application Programming Interface (API) allows third-party tools – such as xiVIEW – to fetch and visualise crosslinking datasets interactively. This expands the potential audience for crosslinking data, making it more accessible not only to structural biologists and computational modellers, but also to machine learning approaches. Integrating a tool like xiVIEW inside the PRIDE structure would then allow crosslinking data submissions to meet the repository's complete submission requirements (as this is dependent on the repository being able to visualise the submitted data, which in addition to mzIdentML also necessitates the peak files to be uploaded). Complete submission status is currently only achievable when submitting crosslink data to JPOST (37). Other ProteomeXchange repositories like MassIVE (38) could similarly incorporate crosslinking-specific workflows, expanding the global interchange for Crosslinking MS data.

While mzIdentML represents a significant milestone in metadata recording for Crosslinking MS, additional information remains essential for ensuring reproducible datasets and analyses. This includes details on sample preparation, analysis procedures, data processing, filtering, and validation. It is also crucial to define a minimal set of manually recorded metadata that must accompany the data, such as the crosslinker used or the species studied, so that reporting remains consistent.

Some aspects of laboratory experimentation must inevitably be recorded manually, but new tools are needed to extract acquisition parameters directly from the mass spectrometric raw data. As these raw data are already included in ProteomeXchange submissions, such extraction could be performed centrally at the point of upload. Data processing details, however, are best tracked during the analysis itself. Doing this manually invites errors, given the multiple steps and numerous parameters involved. Therefore, the field should embrace pipelines and systems that capture metadata internally and allow forwarding it automatically, i.e. with minimal user effort. Software such as Jupyter notebooks (39) provides one possible route, enabling the construction of analysis pipelines while automatically logging parameters and software versions.

Before creating a fresh checklist of requirements specific to Crosslinking MS, it is worth considering practices already established in proteomics. One example is the MIAPE (Minimal Information about a Proteomics Experiment) guidelines, which offer recommendations for minimal metadata reporting in proteomics (40). These guidelines can serve as a starting point and could be adapted to meet the specialised needs of Crosslinking MS.

**Cascading Data to Structural Repositories**

Where Crosslinking MS stands to have profound impact, is in structural biology. The Protein Data Bank (PDB) is the main database for high-confidence 3D structural data. PDB-IHM is a linked database that incorporates integrative and hybrid models (33). The PDB-IHM provides a place to document any experimental restraints used as structural information to build protein models (41). This should include Crosslinking MS data and the derived distance restraints such that structural models can be validated against experimental distance information during upload. Such integration is important for internal validation of the model and also when wanting to reuse the data in future modelling attempts of the same proteins.

To integrate with PDB-IHM, crosslink identifications must be submitted to a field repository in a clearly interpretable format (again pointing to mzIdentML 1.3.0), accompanied by sufficient evidence (scores, decoy matches, FDR thresholds) to convey reliability (42, 43). Crosslinking MS data have become a standard part of integrative modelling workflows, and structural biologists routinely use crosslink-based restraints. With provenance tracking improved, future models will be more transparent and reproducible. This is a declared objective of the wwPDB and also requires action from the side of the structural biology community (42, 43). Firstly, provenance tracking will require modelling tools to work with the same standard data format (possibly mzIdentML) that is deposited in ProteomeXchange. This is not currently possible and requires developers to take action. Secondly, the validation procedures running within PDB-IHM upon model submission must include crosslinking data checks, a first version of such a validation pipeline has been established recently (https://pdb-ihm.org/validation_help.html#dq-crosslinking-ms).

**Ensuring Reliability Through Error Handling**

One of the most pressing technical debates in Crosslinking MS centres on the best strategies for false discovery rate (FDR) estimation (7). Unlike linear peptide searches, crosslinking must match two peptides simultaneously, substantially expanding the search space and increasing the risk of false positives. Target–decoy approaches - often used to gauge error rates in proteomics - have proven more nuanced in Crosslinking MS (18, 23). Additional problems arise from there being multiple levels of consolidation at which results can be interpreted: crosslink-spectra matches; peptide pairs; residue pairs; protein-protein interactions. These reports also demonstrate more appropriate methods, showing that correctly designed target–decoy procedures can address the added complexity (17, 22).

Reaching a consensus on how best to model decoys, from individual spectrum matches up to the level of protein-protein interactions, remains challenging. Researchers often push the boundaries to discover more crosslinks, making it unlikely that the field will converge on a single, uniform strategy for data processing. Rather than attempting to enforce a one-size-fits-all standard, the community must rally around robust testing and benchmarking methods, ensuring that each innovative approach is rigorously validated before entering mainstream use (17, 23, 44). This need for clear benchmarks becomes even more urgent with the growing use of machine learning, which can easily break existing decoy-based error models, for example due to information leakage between training and test data via individual peptides that participate in multiple crosslinked peptide pairs. When this happens, results quantity increases but quality decreases. As the multitude of approaches prevents a uniform error estimation it demands a uniform benchmarking.

Ensuring uniform benchmarking is crucial for improving the quality and reliability of the reported results across the field. It plays a key role in preventing false positives from contaminating large-scale studies or structural models, which could otherwise introduce noise into biological repositories. Before Crosslinking MS results are automatically integrated into PPI databases and subsequently into broader resources like UniProt, the community must collectively work towards higher standards. Establishing a transparent certification process or clear guidelines may be necessary – an important task to address ahead of the Symposium of Structural Proteomics in Milan, Italy, in 2025.

Yet despite these complexities, success stories abound: it has been shown that FDR strategies can be validated - computationally via entrapment databases or with complementary experimental evidence, offering accuracy on par with other proteomic workflows (17, 22, 44). Clarity in FDR estimation and reporting is pivotal for the continued development of Crosslinking MS as a technology, and its automatic integration into biological databases.

**Quantitative Crosslinking MS and Conformational States**

While Crosslinking MS has traditionally been used to detect protein–protein interactions and provide structural restraints, recent developments in quantitative Crosslinking MS highlight its potential to probe protein conformational states. By measuring abundance changes of crosslinked peptides across experimental conditions, quantitation adds a dynamic dimension to structural proteomics. This capacity opens up opportunities for studying allosteric regulation, ligand-induced structural transitions, and the conformational diversity of protein complexes *in vivo*.

These information dimensions elevate the importance of not just detecting crosslinks, but also capturing their quantitative behaviour across experimental conditions. However, most data formats and repositories do not yet support structured representation of quantitative crosslinking information. mzIdentML currently lacks a dedicated schema for encoding crosslink intensities or ratios, and while some repositories such as PRIDE and MassIVE support quantitative proteomics, these functions are not tailored to the requirements of Crosslinking MS. In the short term, extended metadata tables and supplementary annotation files may serve to link crosslinks to experimental conditions. In the long term, standard formats must evolve to include quantitative values and associated experimental metadata, enabling integration of quantitative Crosslinking MS with statistical and modelling tools.

Quantitative Crosslinking MS therefore represents both a challenge and an opportunity. On the one hand, it adds complexity to data annotation and model validation. On the other, it opens new avenues for structural and functional interpretation, including ensemble modelling and the dissection of protein plasticity. Integrating quantitative data into the broader Crosslinking MS ecosystem will require parallel advances in data formats, repositories, and visualization tools that can reflect not just static interactions, but dynamic molecular behaviour.

**Implementing a Federated Repository Ecosystem**

Bringing these threads together, a federated repository ecosystem can position Crosslinking MS data alongside structural and functional information from other approaches. Establishing uniform data submissions and quality control to PRIDE and PDB-IHM ensures that crosslink-derived restraints, PPIs, and structural insights become a valuable component of the global biological knowledge base.

An integrated data ecosystem benefits everyone. Repositories will accept not only raw MS data and peak lists, but also validated crosslink identifications, structural models, annotations, and quality metrics in an easily traceable format that will make them more usable. Journals, funding agencies, and community leaders will encourage or require complete Crosslinking MS data depositions, analogous to the norms established in proteomics and structural biology raising standards for all. Over time, this federated approach lowers barriers for cross-domain research, encourages consistent re-analysis and the development of new tools, and raises overall confidence in crosslink-based findings.

**The Road Ahead**

Crosslinking MS has reached a pivotal juncture. Technical and methodological strides have proved it can offer unique structural insights, but incomplete adoption of data standards and uneven error modelling hold back wider impact. In short, the community now possesses the tools and protocols to remedy these issues – but must act collectively. By (1) embracing mzIdentML 1.3 and robust error models, including reaching consensus on the metadata to be recorded, (2) depositing complete datasets in FAIR-compliant repositories, and (3) building a federated data-sharing framework aligned with structural biology, Crosslinking MS can reach its full potential. Such a future will see crosslinking data routinely displayed in leading databases, cross-referenced with structural and proteomics analyses, and widely trusted as part of the standard toolkit for describing molecular machines. In doing so, we not only bolster the reliability and impact of Crosslinking MS, but also accelerate discoveries in protein function, complex assembly, and systems biology.

**Acknowledgements**

Funded by the Deutsche Forschungsgemeinschaft (DFG, German Research Foundation) under Germany's Excellence Strategy – EXC 2008 – 390540038 – UniSysCat.